\documentstyle[12pt,epsfig]{article}

\newcommand{\be}{\begin{equation}}
\newcommand{\ee}{\end{equation}}

\begin{document}
\topmargin 0pt
\oddsidemargin=0.5truecm
\renewcommand{\thefootnote}{\fnsymbol{footnote}}
\newpage
\setcounter{page}{0}
\begin{titlepage}
\vspace*{-2.0cm}
\begin{flushright}
hep-ph/0602084
\end{flushright}
\vspace*{0.1cm}
\begin{center}
{\Large \bf Neutrino Mass Matrices with Vanishing Determinant
} \\
\vspace{0.6cm}

\vspace{0.4cm}
{\large
Bhag C. Chauhan\footnote{On leave from Govt. Degree College, Karsog (H P)
India 171304. \\
E-mail: chauhan@cftp.ist.utl.pt},
Jo\~{a}o Pulido\footnote{E-mail: pulido@cftp.ist.utl.pt}}\\
\vspace{0.15cm}
{  {\small \sl Centro de F\'{\i}sica Te\'{o}rica das Part\'{\i}culas (CFTP) \\
 Departamento de F\'{\i}sica, Instituto Superior T\'{e}cnico \\
Av. Rovisco Pais, P-1049-001 Lisboa, Portugal}\\
}
\vspace{0.25cm}
and \\
\vspace{0.25cm}
{\large Marco Picariello\footnote{E-mail: Marco.Picariello@le.infn.it}} \\
\vspace{0.15cm}
{\small \sl  I.N.F.N. - Milano, and Dip. di Fisica, Universit\`a di Milano\\
and \\
Dipartimento di Fisica, Universit\`a di Lecce\\
Via Arnesano, ex Collegio Fiorini, I-73100 Lecce, Italia}
\end{center}
\vglue 0.6truecm
\begin{abstract}

We investigate the prospects for neutrinoless double beta decay, texture zeros 
and equalities between neutrino mass matrix elements in scenarios with vanishing 
determinant mass matrices for vanishing and finite $\theta_{13}$ mixing angle in 
normal and inverse mass hierarchies. For normal hierarchy and both zero and finite 
$\theta_{13}$ it is found that neutrinoless double beta decay cannot be observed 
by any of the present or next generation experiments, while for inverse hierarchy 
it is, on the contrary, accessible to experiments. Regarding texture zeros and 
equalities between mass matrix elements, we find that in both normal and inverse 
hierarchies with $\theta_{13}=0$ no texture zeros nor any such equalities can 
exist apart from the obvious ones. For $\theta_{13}\ne 0$ some texture zeros become 
possible. In normal hierarchy two texture zeros occur if $8.1\times 10^{-2}\leq
|sin~\theta_{13}|\leq 9.1\times 10^{-2}$ while  
in inverse hierarchy three are possible, one with $|sin~\theta_{13}|\geq 7\times 10^{-3}$
and two others with $|sin~\theta_{13}|\geq 0.18$. All 
equalities between mass matrix elements are impossible with $\theta_{13}\ne 0$.  

\end{abstract}

\end{titlepage}
\renewcommand{\thefootnote}{\arabic{footnote}}
\setcounter{footnote}{0}

\section{Introduction}
It is a well known fact that the neutrino mass matrix contains nine 
parameters while feasible experiments can hope to determine only seven 
of them in the foreseeable future. This situation can however be overcome,
with the number of free parameters being reduced, if physically 
motivated assumptions are made to restrict the form of the matrix. Among 
the most common such assumptions and as an incomplete list one may refer the
texture zeros \cite{Frampton:2002yf}, hybrid textures \cite{Kaneko:2005yz}, 
traceless condition \cite{He:2003rm}, \cite{He:2003nt}, \cite{Rodejohann:2003ir} 
and vanishing determinant \cite{Branco:2002ie}, \cite{Branco:2003pp},
\cite{Aizawa:2005pk}, the latter two assumptions being basis independent, as 
shall be seen, and the vanishing determinant one equivalent to one vanishing
neutrino mass.

In this paper we perform an investigation on vanishing determinant neutrino
mass matrices aimed at neutrinoless double beta decay ($0\nu\nu\beta\beta$),
texture zeros and equalities
between mass matrix entries. We will assume that neutrinos are
Majorana \cite{Kayser:2005cy}, as favoured by some experimental evidence
\cite{Klapdor-Kleingrothaus:2005zk}, and study the neutrino mass matrix $M$ 
in the weak basis where all charge leptons are already diagonalized. This 
is related to the diagonal mass matrix $D$ through the unitary transformation
\be
D=U^{T}_{MNS}MU_{MNS}
\ee
where we use the standard parametrization \cite{PDG}
\be
U_{MNS}=\!\!\left(\begin{array}{ccc}
c_{12}c_{13} & s_{12}c_{13} & s_{13}e^{-i\delta}\\
-s_{12}c_{23}-c_{12}s_{23}s_{13}e^{i\delta} & c_{12}c_{23}-s_{12}s_{23}s_{13}e^{i\delta}&
s_{23}c_{13} \\
s_{12}s_{23}-c_{12}c_{23}s_{13}e^{i\delta} & -c_{12}s_{23}-s_{12}c_{23}s_{13}e^{i\delta}&
c_{23}c_{13}\\
\end{array}\right).
\ee
where $\delta$ is a Dirac CP violating phase. Equation (1) is equivalent to
\be
M=U^{*}diag(m_{1},m_{2}e^{i\phi_1},m_{3}e^{i\phi_2})U^{\dagger}
\ee
where $\phi_{1},~\phi_{2}$ are two extra CP violating Majorana phases and 
$D=diag(m_{1},m_{2}e^{i\phi_1},m_{3}e^{i\phi_2})$. Applying
determinants properties
\be
\begin{array}{ll}
det~M&=det~(U^{*}DU^{\dagger}) \\
&=det~(U^{*}U^{\dagger}D) \\
&=det~U^{*}\ det~U^{\dagger}\ det~D \\
&=det~D~~(U~real) \\
&\ne det~D~~(U~complex)
\end{array}
\ee
because if matrix $U$ is real, $U^{*}U^{\dagger}=UU^{T}=1$, which is satisfied provided 
$\delta=0$ or $\theta_{13}=0$ (see eq.(2)).
Thus the determinant is not in general basis independent. In order
that $det~D=det~M$ it is necessary and sufficient that there is either no Dirac CP violation
or that it is unobservable. 
The same arguments hold for the condition $TrD=TrM$ \cite{He:2003nt}.

From eq. (4) we get that $det~M=0$ if and only if $det~D=0$,
because $det U^{\dagger}$ and $det U^{*}$ are not zero.
The vanishing determinant condition is basis independent, 
corresponding to a zero eigenvalue of the mass matrix.
So requiring $det~M=0$ is equivalent to assuming one of the neutrinos
to be massless. This is realized for instance in 
the Affleck-Dine scenario for leptogenesis \cite{Affleck:1984fy},\cite{Murayama:1993em},
\cite{Dine:1995kz} which requires the lightest neutrino to be practically massless
($m\simeq 10^{-10} eV$) \cite{Asaka:2000nb},\cite{Fujii:2001sn}. We will 
consider separately the cases of vanishing and finite $\theta_{13}$
\footnote{The $2\sigma$ range recently
obtained for this quantity is \cite{Fogli:2005cq} $sin^2 \theta_{13}=
0.9\pm^{2.3}_{0.9}\times 10^{-2}$, the lower uncertainty being purely formal,
corresponding to the positivity constraint $sin^{2} \theta_{13}\geq 0$.}. 
In the first the Dirac phase is unobservable and
the usual definition $U_{MNS}=U_{23}U_{13}U_{12}$ 
\cite{King:2002gx} simplifies to $U_{MNS}=U_{23}U_{12}$ with
\be
U_{23}=\left(\begin{array}{ccc} 1  &  0  &  0 \\
0   &  \alpha_{22}  &    \alpha_{23}   \\
0   &  \alpha_{32}  &    \alpha_{33}    \end{array}\right),
~U_{12}=\left(\begin{array}{ccc} \beta_{11}  &  \beta_{12}  &  0 \\
\beta_{21}   &  \beta_{22}  &    0   \\
0   &    0  &    1  \\  \end{array}\right) 
\ee
where the unitarity condition $(|\alpha_{22}\alpha_{33}-\alpha_{32}\alpha_{23}|=
|\beta_{11}\beta_{22}-\beta_{12}\beta_{21}|=1)$
implies $\alpha_{22}\alpha_{33}\alpha_{32}\alpha_{23}<0$ and 
$\beta_{11}\beta_{22}\beta_{12}\beta_{21}<0$ with
$\alpha_{22}=\pm cos \theta_{\otimes}$, $\beta_{11}=\pm cos \theta_{\odot}$,
the remaining matrix elements being evident. For neutrino masses and
mixings we refer to the following $2\sigma$
ranges~\cite{Fogli:2005cq,Gonzalez-Garcia:2004jd}
\be
\Delta m^2_{\odot}=m^2_{2}-m^2_{1}=7.92 \times 10^{-5}(1\pm 0.09) eV^2,
\ee
\be
\Delta m^2_{\otimes}=m^2_{3}-m^2_{2}=\pm 2.4 \times 10^{-3}(1\pm^{0.21}_{0.61}) eV^2
\ee
\be
sin^2\theta_{\odot}=0.314(1\pm^{0.18}_{0.15}),
\ee
\be
sin^2\theta_{\otimes}=0.44(1\pm^{0.41}_{0.22})
\ee
obtained from a 3 flavour analysis of all solar and atmospheric data. This favours
the widely used form of the $U_{MNS}$ matrix \cite{Scott:1999bs} (all entries taken 
in their moduli)
\be
U_{MNS}=\left(\begin{array}{ccc} \sqrt{\frac{2}{3}}&   \frac{1}{\sqrt{3}} & 0 \\
\frac{1}{\sqrt{6}} &   \frac{1}{\sqrt{3}} & \frac{1}{\sqrt{2}} \\
\frac{1}{\sqrt{6}} &   \frac{1}{\sqrt{3}} & \frac{1}{\sqrt{2}} \\
\end{array}\right).
\ee
For finite $\theta_{13}$ we will work in the approximation \cite{Fogli:2005cq}
\be
U_{13}=\left(\begin{array}{ccc} \gamma_{11} & 0 & \gamma_{13} \\
0 & 1 & 0 \\ \gamma_{31} & 0 & \gamma_{33}\end{array}\right) \simeq 
\left(\begin{array}{ccc} 1 & 0 & \gamma_{13} \\ 0 & 1 & 0 \\
\gamma_{31} & 0 & 1 \end{array}\right)
\ee
with $\gamma_{13}=U_{e3}=s_{13}e^{i\delta}=-\gamma_{31}^{*}$. This leads to 
\be
U_{MNS}=\left(\begin{array}{ccc} \beta_{11} & \beta_{12} & \gamma_{13} \\
\alpha_{23}\gamma_{31}\beta_{11}+\alpha_{22}\beta_{21} &
\alpha_{23}\gamma_{31}\beta_{12}+\alpha_{22}\beta_{22} & \alpha_{23} \\
\alpha_{33}\gamma_{31}\beta_{11}+\alpha_{32}\beta_{21} &
\alpha_{33}\gamma_{31}\beta_{12}+\alpha_{32}\beta_{22} & \alpha_{33}
\end{array}\right)
\ee
which generalizes eq.(10) for small $s_{13}$.

The paper is organized as follows: in section 2 we derive all possible forms
of the mass matrix $M$ in this scenario for both normal and inverse 
hierarchies and investigate their consequences for $0\nu\nu\beta\beta$ decay. 
Since one of the neutrinos is massless, there is only one Majorana phase to 
be considered. In section 3 we investigate the prospects for texture zeros 
and equalities between matrix elements in both hierarchies. In sections 2 and 3
only $\theta_{13}=0$ is considered. In section 4 the previous analysis is 
extended to $\theta_{13}\ne 0$ and in section 5 we briefly expound our main 
conclusions.

\section{Mass matrices with vanishing $\theta_{13}$: $\beta \beta_{0\nu\nu}$ decay}

\subsection{Normal hierarchy (NH)}  

This is the case where the two mass eigenstates involved in the solar oscillations
are assumed to be the lightest so that $\Delta m^2_{\otimes}=\Delta m^2_{32}>0$.
We will consider this case as a departure from the degenerate one with 
$\Delta m^2_{\odot}=\Delta m^2_{21}=0$ and break the degeneracy with a real 
parameter $\epsilon$. Matrix $D$ with $m$ and $\epsilon$ both real is therefore
\be
D=diag(0,3 \epsilon e^{i\phi},m)
\ee
where $\phi$ is the Majorana relative phase between the second and third diagonal 
matrix elements ($\phi=\phi_1-\phi_2$ in the notation of section 2) and 
$\Delta m^2_{\odot}=9 \epsilon^2$. Using eqs.(5) the matrix $M$ is
\be
M\!\!=\!\!U_{23}U_{12}DU_{12}^T U_{23}^T\!\!=\!\!\left(\begin{array}{ccc}
3\epsilon e^{i\phi} \beta^2_{12} & 3\epsilon e^{i\phi} \alpha_{22}\beta_{12}\beta_{22} &
3\epsilon e^{i\phi} \alpha_{32}\beta_{12}\beta_{22} \\
3\epsilon e^{i\phi} \alpha_{22}\beta_{12}\beta_{22} &
3\epsilon e^{i\phi} \alpha_{22}^2 \beta_{22}^2 \!\!+\!\!m\alpha_{23}^2 &
3\epsilon e^{i\phi} \alpha_{22} \alpha_{32} \beta_{22}^2 \!\!+\!\!m\alpha_{23} \alpha_{33} \\
3\epsilon e^{i\phi} \alpha_{32}\beta_{12}\beta_{22} &
3\epsilon e^{i\phi} \alpha_{22} \alpha_{32} \beta_{22}^2 \!\!+\!\!m\alpha_{23} \alpha_{33} &
3\epsilon e^{i\phi} \alpha_{32}^2 \beta_{22}^2 \!\!+\!\!m\alpha_{33}^2 \\ 
\end{array}\right).
\ee
Owing to the sign ambiguities of parameters $\alpha$ and $\beta$, four possibilities
for matrix $M$ arise. 
Suppose entries 12 and 13 in this matrix have (+) (+) signs. Then $\alpha_{22},
\alpha_{32}$ have the same sign as $\beta_{12} \beta_{22}$, that is $\alpha_{22}
\alpha_{32}$ in the (23) entry is (+), implying the opposite sign for the
coefficient of $m$ ($\alpha_{23}\alpha_{33}$). 
So eq.(12) has the form 
\be
M=\left(\begin{array}{ccc} \epsilon e^{i\phi} &   \epsilon e^{i\phi} & \epsilon  e^{i\phi}\\
\epsilon e^{i\phi} & (m/2)+\epsilon e^{i\phi} &  -(m/2)+\epsilon  e^{i\phi}\\
\epsilon e^{i\phi} & -(m/2)+\epsilon e^{i\phi} &  (m/2)+\epsilon  e^{i\phi}\\ \end{array}\right)
\ee
Suppose entries 12 and 13 in the matrix have (-) (-) signs. Then $\alpha_{22},
\alpha_{32}$ have opposite sign to $\beta_{12} \beta_{22}$, that is they have
the same sign, so $\alpha_{22}\alpha_{32}$ is (+) and $\alpha_{23}\alpha_{33}$
is (-) so
\be
M=\left(\begin{array}{ccc} \epsilon  e^{i\phi}&   -\epsilon e^{i\phi} & -\epsilon  e^{i\phi}\\
-\epsilon e^{i\phi} & (m/2)+\epsilon  e^{i\phi}&  -(m/2)+\epsilon e^{i\phi} \\
-\epsilon e^{i\phi} & -(m/2)+\epsilon e^{i\phi} &  (m/2)+\epsilon e^{i\phi} \\ \end{array}\right)
\ee
Suppose entries 12 and 13 in the matrix have (+) (-) signs. Then $\alpha_{22},
\alpha_{32}$ have opposite signs to each other, so $\alpha_{22}\alpha_{32}$ is (-)
and $\alpha_{23}\alpha_{33}$ is (+). Hence
\be
M=\left(\begin{array}{ccc} \epsilon e^{i\phi} &   \epsilon e^{i\phi} & -\epsilon e^{i\phi} \\
\epsilon e^{i\phi} & (m/2)+\epsilon e^{i\phi} &  (m/2)-\epsilon e^{i\phi} \\
-\epsilon e^{i\phi} & (m/2)-\epsilon e^{i\phi} &  (m/2)+\epsilon  e^{i\phi}\\ \end{array}\right)
\ee
Suppose entries 12 and 13 in the matrix have (-) (+) signs. Then $\alpha_{22},
\alpha_{32}$ have opposite signs to each other, so $\alpha_{22}\alpha_{32}$ is (-)
and $\alpha_{23}\alpha_{33}$ is (+). Hence the matrix is
\be
M=\left(\begin{array}{ccc} \epsilon e^{i\phi} &   -\epsilon e^{i\phi} & \epsilon  e^{i\phi}\\
-\epsilon  e^{i\phi}& (m/2)+\epsilon  e^{i\phi}&  (m/2)-\epsilon  e^{i\phi}\\
\epsilon e^{i\phi} & (m/2)-\epsilon  e^{i\phi}&  (m/2)+\epsilon  e^{i\phi}\\ \end{array}\right)
\ee
All matrices (15), (16), (17), (18) have vanishing determinant as can be easily verified. 
For $0\nu\nu\beta\beta$ decay
\be
<m_{ee}>=|U_{e1}^2m_{1}+U_{e2}^2m_{2}e^{i\phi_1}+U_{e3}^2m_{3}e^{i\phi_2}| 
\ee
hence, for $m_1=0$ and $U_{e3}=\gamma_{13}=s_{13}e^{i\delta}=0$
\be
<m_{ee}>=|U_{e2}^2m_{2}e^{i\phi_1}|=\frac{1}{3}3\epsilon=
\frac{1}{3}\sqrt{\Delta m^2_{\odot}}\simeq 3\times 10^{-3}eV
\ee
where we used $\epsilon=\frac{1}{3}\sqrt{\Delta m^2_{\odot}}$. So the
Majorana phase is not an observable. 

There is no commonly accepted evidence in favour of $0\nu\nu\beta\beta$ decay  
but there exist reliable upper limits on $<m_{ee}>$ 
\be
<m_{ee}>\leq (0.3-1.2)eV~\cite{Klapdor-Kleingrothaus:2005zk},~~
<m_{ee}>\leq (0.2-1.1)eV~\cite{Arnaboldi:2005cg}
\ee
where the uncertainties follow from the uncertainties in the nuclear matrix
elements. The future CUORE experiment \cite{Capelli:2005jf}, of which 
CUORICINO is a test version \cite{Arnaboldi:2005cg}, is expected to improve
this upper bound to $3\times 10^{-2}eV$. Other experiments are also proposed 
(MAJORANA \cite{Aalseth:2005mn}, GENIUS \cite{Klapdor-Kleingrothaus:1999hk}, 
GEM \cite{Zdesenko:2001ee} and others) in which the sensitivity of a few 
$10^{-2}eV$ is planned to be reached.

Conclusion: {\it vanishing determinant with vanishing $\theta_{13}$ and
NH implies that $0\nu\nu\beta\beta$ decay cannot be detected 
even in the next generation of experiments.} 

\subsection{Inverse Hierarchy (IH)} 

We start with matrix $D$ in the form $D=diag\{m,(m+\epsilon)e^{i\phi},0\}$
where $m$, $\epsilon$ are complex, $|m|\simeq \sqrt{\Delta m^{2}_{\otimes}}$,
$|\epsilon| \simeq \sqrt{\Delta m^{2}_{\odot}}$ and chosen in such
a way that $m+\epsilon=\tilde m$ is real  
($\epsilon=0$ corresponds to the degenerate case). Alternatively
$D=diag\{\tilde m- \epsilon, \tilde m e^{i\phi},0\}$
with, of course,
$\tilde m- \epsilon$ complex. Multiplying the whole matrix by the inverse 
phase of $\tilde m- \epsilon$, it can be redefined as 
\be
D=diag\{\tilde m- \lambda, \tilde m e^{i(\phi-\psi)},0\}
\ee
with $\lambda$ real
and defined by $(\tilde m- \epsilon)e^{-i\psi}= \tilde m- \lambda$ (notice that 
$\tilde m- \epsilon=|\tilde m- \epsilon|e^{i\psi}$ and $\tilde m =
\sqrt{\Delta m^{2}_{\otimes}}$). There are two solutions for $\lambda$. In fact,
imposing the solar mass square difference
\be 
\Delta m^{2}_{\odot}=|d_{22}|^2-|d_{11}|^2=\tilde m^2- \tilde m^2 +2\lambda \tilde {m}-
\lambda^2
\ee
and solving the quadratic equation $\lambda^2 - 2\lambda \tilde m +
\Delta m^{2}_{\odot}=0$ one gets
\be
\lambda=\tilde m\pm\sqrt{\tilde m^2- \Delta m^{2}_{\odot}}=\lambda_{\pm}.
\ee
Notice that $\lambda_{+}$ is large and $\lambda_{-}$ is small. To first order 
in $\frac{\Delta m^2_{\odot}}{\tilde m^2}=\frac{\Delta m^2_{\odot}}
{\Delta m^2_{\otimes}}\simeq 0.30$ one has
\be
\lambda_{+}=\tilde m (2 -\frac{1}{2}\frac{\Delta m^2_{\odot}}{\Delta m^2_{\otimes}})
\simeq 1.85 \tilde m
\ee
\be
\lambda_{-}=\frac{\tilde m}{2}\frac{\Delta m^2_{\odot}}{\Delta m^2_{\otimes}}
\simeq \frac{\tilde m}{60}
\ee
It is straightforward to see that $D(\lambda_{-},\phi+ \pi)=-D(\lambda_{+},\phi)$
and the same property holds for matrix $M$, namely $M(\lambda_{-},\phi+ \pi)=
-M(\lambda_{+},\phi)$ because $U_{MNS}$ is invariant under the transformations
$\lambda_{+} \rightarrow \lambda_{-}$ and $\phi \rightarrow \phi +\pi$. 
So the two solutions for $\lambda$ are equivalent: one may take either
\be 
\lambda_{+}~,~\psi=0
\ee
or
\be
\lambda_{-}~,~\psi=\pi.
\ee
Using $M\!=\!U_{23}U_{12}DU_{12}^T U_{23}^T$ with eqs.(5), (22), the matrix $M$ 
has now the form
\be
M=\left(\begin{array}{ccc}
\tilde m(1-\frac{t}{3})-\frac{2}{3}\lambda & (sign)\frac{1}{3}(\tilde {m}t-\lambda) & 
(sign)\frac{1}{3}(\tilde {m}t-\lambda) \\
(sign)\frac{1}{3}(\tilde {m}t-\lambda) & \tilde m(\frac{1}{2}-\frac{t}{3})-\frac{\lambda}{6} &
(sign)[\tilde m(\frac{1}{2}-\frac{t}{3})-\frac{\lambda}{6}] \\
(sign)\frac{1}{3}(\tilde {m}t-\lambda) & (sign)[\tilde m(\frac{1}{2}-\frac{t}{3})-\frac{\lambda}{6}] &
\tilde m(\frac{1}{2}-\frac{t}{3})-\frac{\lambda}{6} \\
\end{array}\right) 
\ee
\begin{figure}[h]
\setlength{\unitlength}{1cm}
\begin{center}
\hspace*{-1.6cm}
\epsfig{file=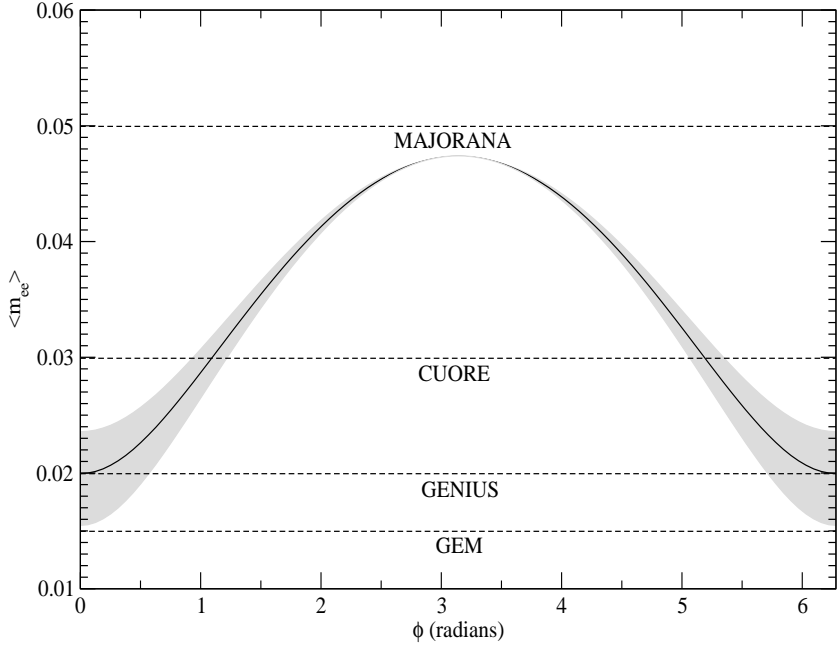,height=13.0cm,width=11.0cm,angle=270}
\end{center}
\caption{ \it$\beta \beta_{0\nu\nu}$ decay effective mass parameter $<m_{ee}>$ as a function
of the Majorana phase $\phi$ showing its accessibility for forthcoming experiments.}
\label{fig1}
\end{figure}
which also verifies $det~M=0$ as expected.
Equation (29) is formally the same for $\lambda=\lambda_{+}$ and $\lambda=\lambda_{-}$
with the definition $t=1-e^{i\phi}$ for $\lambda=\lambda_{+}$ and $t=1+e^{i\phi}$
for $\lambda=\lambda_{-}$, the sign affecting the exponential 
being related to the $\psi$ phase.
The structure of (+) and (-) in eq.(29) is the same as before ((15), (16), (17), (18)):
equal signs in entries $M_{12}$, $M_{13}$ correspond to (+) in both entries $M_{23}$, 
$M_{32}$ while different signs in $M_{12}$, $M_{13}$ correspond to (-) in both entries 
$M_{23}$, $M_{32}$.
Eq. (29) is the equivalent for IH of (15), (16), (17), (18) for NH. 

For $\beta \beta_{0\nu\nu}$ decay we have
\be
<m_{ee}>=|U_{e1}^2m_{1}+U_{e2}^2m_{2}e^{i\phi_1}+U_{e3}^2m_{3}e^{i\phi_2}|=
|\frac{2}{3}(\tilde m- \lambda_{\pm})\pm \frac{1}{3}\tilde m e^{i(\phi)}|. 
\ee
The quantity $m_{ee}$ is displayed in fig.1 as a function of the phase difference $\phi$.
The shaded areas correspond to the $1\sigma$ uncertainties in the solar angle $\theta_{\odot}$.
It is seen from eq.(30) and fig.1 that for inverse hierarchy (vanishing $\theta_{13}$ 
and mass matrix determinant) $\beta \beta_{0\nu\nu}$ decay is phase dependent and within 
observational limits of forthcoming experiments. So:

Conclusion: {\it models with vanishing determinant mass matrix and vanishing $\theta_{13}$
provide, in inverse hierarchy, a Majorana phase dependent $\beta \beta_{0\nu\nu}$ decay 
which is physically observable for most values of the phase 
in the next generation of experiments.}


\section{Texture zeros and equalities between M matrix elements ($\theta_{13}=0$)}

\subsection{Texture zeros}

Here we analyze the possibility of vanishing entries in the mass matrix $M$. 
Taking first NH and recalling eqs.(15)-(18), it is seen that this implies
either $m /2=\pm \epsilon e^{i \phi}$ or $\epsilon=0$, both situations 
being impossible. For IH three cases need to be considered: \\
$(a)~~M_{11}=0$ \\
We have in this case $\tilde m(1-\frac{t}{3})-\frac{2}{3}\lambda=0$ implying
\be
\tilde m(3-t)=2\lambda.
\ee
Replacing $t\rightarrow 1-e^{i\phi}$ and $\lambda \rightarrow \lambda_{+}$ this
leads to
\be
e^{i\phi}=2\sqrt{1-\frac{\Delta m^2_{\odot}}{\Delta m^2_{\otimes}}}
\ee
\vspace{0.3cm}
which is experimentally excluded. \\   
$(b)~~M_{12}=0$ \\
This gives $\tilde m t-\lambda=0$, hence using the same replacement
\be
e^{i\phi}=-\sqrt{1-\frac{\Delta m^2_{\odot}}{\Delta m^2_{\otimes}}}
\ee
\vspace{0.3cm}
which is also impossible since $\Delta m^2_{\odot}=0$ is strictly excluded experimentally. \\
$(c)~~M_{22}=0$ \\ 
This gives $\tilde m(\frac{1}{2}-\frac{t}{3})-\frac{\lambda}{6}=0$, hence using the same 
replacement 
\be
e^{i\phi}=\frac{1}{2}\sqrt{1-\frac{\Delta m^2_{\odot}}{\Delta m^2_{\otimes}}}
\ee
which is also experimentally excluded. In the former cases (a), (b), (c) the same 
results are of course obtained with the replacement $t\rightarrow 1+e^{i\phi}$ and 
$\lambda \rightarrow \lambda_{-}$, as can be easily verified. 
So zero mass textures are not possible in the present scenario. 

The same conclusion can be obtained using the results from the literature. In fact
the analytical study of various structures of the neutrino mass matrix was
presented systematically by Frigerio and Smirnov \cite{Frigerio:2002fb} who also discussed 
the case of equalities of matrix elements. Here we use a result from \cite{Xing:2003ic} 
where specific relations among the mixing angles were derived for one texture zero and 
one vanishing eigenvalue. We refer to table I of \cite{Xing:2003ic} and first to NH. 
Using their 
definition of parameter $\chi=\left|\frac{m_2}{m_3}\right|$ we have in our model $\chi=
\sqrt {\frac{\Delta m^2_{\odot}}{\Delta m^2_{\otimes}}} =0.182$ and so for 
cases A, B, C, D, E, F respectively in their notation
\be
\chi=0,~\chi=0,~\chi=0,~\chi=1.50,~\chi=1.50,~\chi=1.50 
\ee
For inverse hierarchy, defining  
$\eta=\frac{m_1}{m_2}=\frac{|\tilde m -\lambda_{\pm}|}{\sqrt{\Delta m^2_{\otimes}}}=
\sqrt{1-\frac{\Delta m^2_{\odot}}{\Delta m^2_{\otimes}}}=0.983$ we have for cases 
A, B, C, D, E, F respectively
\be
\eta=0.50,~\eta=1,~\eta=1,~\eta=2.0,~\eta=2.0,~\eta=2.0
\ee
Notice that $0.953<\eta < 0.988$ (using $1\sigma$ upper and lower values for the solar 
and atmospheric mass square differences). So one can draw the following: 

\vspace{0.6cm}
{\it Conclusion: both NH and IH cannot work with det D= det M=0, vanishing $\theta_{13}$
and one texture zero. In other words, vanishing determinant scenarios with $\theta_{13}=0$
are experimentally excluded, unless they have no texture zeros}.

\subsection{Equalities between matrix elements}

First we consider the case of NH. Equations (15)-(18) can be written in the general form
\be
M=\left(\begin{array}{ccc} \epsilon e^{i\phi} &  sign(\epsilon e^{i\phi}) & sign(\epsilon  e^{i\phi})\\
sign(\epsilon e^{i\phi}) & (m/2)+\epsilon e^{i\phi} & sign[-(m/2)+\epsilon e^{i\phi}]\\
sign(\epsilon e^{i\phi}) & sign[-(m/2)+\epsilon e^{i\phi}] &  (m/2)+\epsilon  e^{i\phi}\\ \end{array}\right)
\ee
and using the same sign conventions as in eqs.(15)-(18), it is seen that $|M_{11}|=|M_{12}|=|M_{13}|$ 
and $M_{22}=M_{33}$. The first three equalities provide either $\epsilon=0$, which is
excluded by data, or identities. Hence the
relations to be investigated are $M_{11}=M_{22}$, $M_{11}=M_{23}$, $M_{22}=M_{23}$.

Equation $M_{11}=M_{22}$ implies $m=0$ which is impossible.

Equation $M_{11}=M_{23}$ yields either
\be (a)~~\epsilon e^{i\phi}=(-m/2)+\epsilon e^{i\phi} 
\ee
leading to $m=0$, or
\be (b)~~\epsilon e^{i\phi}=(+m/2)-\epsilon e^{i\phi} 
\ee
leading to $\frac{4}{3}=\sqrt{1+\frac{\Delta m^2_{\otimes}}{\Delta m^2_{\odot}}}$
which is also experimentally excluded.

Equation $M_{22}=M_{23}$ yields either
\be (a)~~(+m/2)+\epsilon e^{i\phi}=(-m/2)+\epsilon e^{i\phi}
\ee
leading to $m=0$, or
\be (b)~~(+m/2)+\epsilon e^{i\phi}=(+m/2)-\epsilon e^{i\phi}
\ee
leading to $\epsilon=0$, both experimentally excluded.

Next we consider IH. We use eq. (29) and note that the matrix is symmetric, so there are
at first sight 6 independent entries. However $M_{22}=M_{33}$, $|M_{12}|=|M_{13}|$,
$|M_{22}|=|M_{23}|$. So apart from the obvious ones, there are three equalities to be
investigated: $M_{11}=M_{12}$, $M_{11}=M_{23}$, $M_{12}=M_{23}$.

Equality $M_{11}=M_{12}$ yields two cases
\be (a)~~\tilde m(1-\frac{t}{3})-\frac{2}{3}\lambda=\frac{1}{3}(\tilde m t-\lambda)
\ee
which upon using $\lambda=\lambda_{\pm}$ for $t=1\mp e^{i\phi}$ gives  
\be
\tilde m -\lambda_{\pm}={\mp}2 \tilde m e^{i\phi}
\ee
which is impossible to satisfy, as seen from eq.(22), and 
\be
(b)~~\tilde m(1-\frac{t}{3})-\frac{2}{3}\lambda=-\frac{1}{3}(\tilde m t-\lambda)
\ee
leading to 
\be
\tilde m=\lambda,
\ee
also impossible, eq.(24).

Equality $M_{11}=M_{23}$. The two cases to be considered are
\be
(a)~~\tilde m(1-\frac{t}{3})-\frac{2}{3}\lambda= \tilde m(\frac{1}{2}-\frac{t}{3})-\frac{\lambda}{6}
\ee
from which
\be
\tilde m=\lambda 
\ee
which cannot be satisfied (eq.(24)) and
\be
(b)~~\tilde m(1-\frac{t}{3})-\frac{2}{3}\lambda=-\tilde m(\frac{1}{2}-\frac{t}{3})+\frac{\lambda}{6}
\ee
which upon using $\lambda=\lambda_{\pm}$ for $t=1\mp e^{i\phi}$ gives
\be
\tilde m -\lambda_{\pm}={\mp}\frac{4}{5}\tilde m e^{i\phi}
\ee
or equivalently
\be
5\sqrt{1-\frac{\Delta m^2_{\odot}}{\Delta m^2_{\otimes}}}=4e^{i\phi}
\ee 
which is cannot be satisfied even if $\phi=0$. (Maximizing $\Delta m^2_{\odot}$
and minimizing $\Delta m^2_{\otimes}$ (1 $\sigma$) the above square root verifies 
$0.953<\sqrt{1-\frac{\Delta m^2_{\odot}}{\Delta m^2_{\otimes}}}<0.988$).

Equality $M_{12}=M_{23}$. The two cases are now
\be
(a)~~\frac{1}{3}(\tilde m t-\lambda)=\tilde m(\frac{1}{2}-\frac{t}{3})-\frac{\lambda}{6}
\ee
which gives $\tilde m-\lambda_{\pm}=\pm 4\tilde m e^{i\phi}$ or $\pm \sqrt{1-\frac{\Delta m^2_{\odot}}
{\Delta m^2_{\otimes}}}=\pm 4e^{i\phi}$, again impossible, and
\be
(b)~~\frac{1}{3}(\tilde m t-\lambda)=-\tilde m(\frac{1}{2}-\frac{t}{3})-\frac{\lambda}{6}
\ee
or $\tilde m=\lambda$, also impossible.
All these impossibilities mean {\it experimentally excluded}.

Moreover, it is seen from eq.(29) that if $M_{12}$ and $M_{13}$ have opposite signs,
requiring their equality implies they both vanish, leading to  
two texture zeros which is excluded. The same is true for $M_{22}$ and 
$M_{23}$. Recall that one texture zero with vanishing determinant cannot work with 
$\theta_{13}=0$ (see section 3.1). Hence:

\vspace{0.6cm}
{\it Conclusion: equalities between mass matrix elements with $\theta_{13}=0$ apart from 
the obvious ones are experimentally excluded.}

\section{Mass matrices with $\theta_{13}\ne 0$} 

Regarding $\beta\beta_{0\nu\nu}$ decay, it can easily be seen (eq.(19)) that the conclusions 
derived in section 2 for $\theta_{13}=0$ remain unchanged here both for normal and inverse
hierarchies.

The form of matrix $U$ can now be derived from eqs.(5), (11). We have 
\be
U=U_{23}U_{13}U_{12}=U_{12}^TU_{13}^TU_{23}^T.
\ee
For normal and inverse hierarchies, using the notation of sections 1, 2, we have for the mass 
matrix $M$ and recalling that $M$ is symmetric 

{\it Normal Hierarchy} 
\be
\begin{array}{lll} 
M_{11}\!&\!=\!&\!3\epsilon \beta_{12}^2e^{i\phi}+m\gamma_{13}^2 \\
M_{12}\!&\!=\!&\!3\epsilon(\alpha_{23}\gamma_{31}\beta_{12}^2+\alpha_{22}\beta_{22}\beta_{12})e^{i\phi}
+m\alpha_{23}\gamma_{13} \\
M_{13}\!&\!=\!&\!3\epsilon(\alpha_{33}\lambda_{31}\beta_{12}^2+\alpha_{32}\beta_{22}\beta_{12})e^{i\phi}
+m\alpha_{33}\gamma_{13} \\
M_{22}\!&\!=\!&\!3\epsilon(\alpha_{23}\gamma_{31}\beta_{12}+\alpha_{22}\beta_{22})e^{i\phi}
+m\alpha_{23}^2 \\
M_{23}\!&\!=\!&\!3\epsilon(\alpha_{23}\gamma_{31}\beta_{12}+\alpha_{22}\beta_{22})
(\alpha_{33}\gamma_{31}\beta_{12}+\alpha_{32}\beta_{22})e^{i\phi}+m\alpha_{23}\alpha_{33} \\
M_{33}\!&\!=\!&\!3\epsilon(\alpha_{33}\gamma_{31}\beta_{12}+\alpha_{32}\beta_{22})e^{i\phi}
+m\alpha_{33}^2 \\
\end{array}
\ee

{\it Inverse Hierarchy}
\be
\begin{array}{lll} 
M_{11}\!\!&\!\!\!=\!\!\!&\!\!\!(\tilde m-\lambda)
\beta_{11}^2+\tilde me^{i(\phi-\psi)}\beta_{12}^2 \\
M_{12}\!\!&\!\!\!=\!\!\!&\!\!\!(\tilde m-\lambda)
\beta_{11}(\alpha_{23}\gamma_{31}\beta_{11}+\alpha_{22}\beta_{21})
+\tilde me^{i(\phi-\psi)}\beta_{12}(\alpha_{23}\gamma_{31}\beta_{12}+\alpha_{22}\beta_{22}) \\
M_{13}\!\!&\!\!\!=\!\!\!&\!\!\!(\tilde m-\lambda)
\beta_{11}(\alpha_{33}\gamma_{31}\beta_{11}+\alpha_{32}\beta_{21})
+\tilde me^{i(\phi-\psi)}\beta_{12}(\alpha_{33}\gamma_{31}\beta_{12}+\alpha_{32}\beta_{22}) \\
M_{22}\!\!&\!\!\!=\!\!\!&\!\!\!(\tilde m-\lambda)
(\alpha_{23}\gamma_{31}\beta_{11}+\alpha_{22}\beta_{21})^2
+\tilde me^{i(\phi-\psi)}(\alpha_{23}\gamma_{31}\beta_{12}+\alpha_{22}\beta_{22})^2 \\
M_{23}\!\!&\!\!\!=\!\!\!&\!\!\!(\tilde m-\lambda)\!
(\alpha_{23}\gamma_{31}\beta_{11}\!\!+\!\!\alpha_{22}\beta_{21}\!)
(\alpha_{23}\gamma_{31}\beta_{11}\!\!+\!\!\alpha_{22}\beta_{21}\!)\!+\!\tilde me^{i(\phi-\psi)}
(\alpha_{23}\gamma_{31}\beta_{12}\!\!+\!\!\alpha_{22}\beta_{22})
(\alpha_{33}\gamma_{31}\beta_{12}\!\!+\!\!\alpha_{32}\beta_{22}) \\
M_{33}\!\!&\!\!\!=\!\!\!&\!\!\!(\tilde m-\lambda)
(\alpha_{33}\gamma_{31}\beta_{11}+\alpha_{32}\beta_{21})^2
+\tilde me^{i(\phi-\psi)}(\alpha_{33}\gamma_{31}\beta_{12}+\alpha_{32}\beta_{22})^2 \\
\end{array}
\ee

Starting with the analysis of texture zeros, we use eqs.(54) and require
$|s_{13}|\leq 0.16$ (90\% CL) or $|s_{13}|\leq 0.22$ (99.73\% CL). The NH case
is sufficiently simple to be solved analytically. We investigate in turn each of the 
6 independent equations of 
\vspace{0.4cm}
the form $M_{ij}=0$ for NH. \\
(a) $M_{11}=0$ \\
From the first of eqs.(54), inserting $\gamma_{13}=s_{13}e^{i\delta}$, 
$\beta_{12}^2=1/3$ (see eqs.(10)-(12)) this leads to
\be
s_{13}^2m e^{2i\delta}+\epsilon e^{i\phi}=0
\ee
from which 
\be
\delta=(\phi\pm \pi)/2,~s_{13}=\pm \frac{\epsilon}{m}.
\ee
\begin{figure}[h]
\setlength{\unitlength}{1cm}
\begin{center}
\hspace*{-1.6cm}
\epsfig{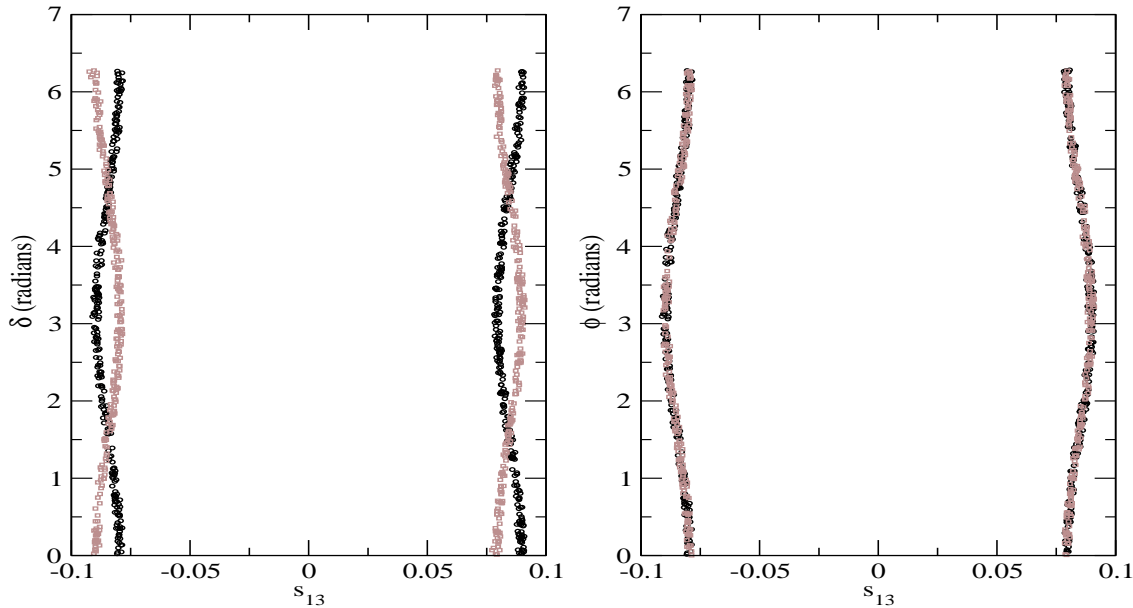}
\end{center}
\caption{ \it Texture zeros in NH: Dirac phase (left panel) and Majorana phase 
(right panel) vs. $s_{13}$ for solutions to $M_{12}=0$ and $M_{13}=0$. Black and 
grey areas represent the two different sets of parameters that satisfy both textures.}
\label{fig2}
\end{figure}
The last relation gives $|s_{13}|=0.25$. The $3\sigma$ upper bound on $|s_{13}|$ is
0.22, so $M_{11}=0$ 
is strongly disfavoured. \\
(b) $M_{12}=0$ \\
Here, as in the following four cases, sign ambiguities must be taken into account.
One gets
\be
\pm \epsilon e^{i\phi}=\pm \alpha_{23}s_{13}m\left(\pm e^{i\delta}\pm\frac{\epsilon}{m}
e^{-i(\delta-\phi)}\right)
\ee
where the signs are uncorrelated and a condition for $s_{13}$ follows
\be
s_{13}=\frac{\epsilon}{\alpha_{23}m}\left(1\pm O(\frac{\epsilon}{m})\right).
\ee 
(c) $M_{13}=0$ \\
In the same way one gets an equation like (58) with the replacement $\alpha_{23}\rightarrow
\alpha_{33}$ and
\be
s_{13}=\frac{\epsilon}{\alpha_{33}m}\left(1\pm O(\frac{\epsilon}{m})\right).
\ee
Hence both $M_{12}=0$ and $M_{13}=0$ 
are allowed with $|s_{13}|$ oscillating around the average 
$\left|\frac{\epsilon}{\alpha_{23}m}\right|=\left|\frac{\epsilon}{\alpha_{33}m}
\right|\simeq 0.086$ and in the interval 
$8.1\times 10^{-2}\leq |s_{13}|\leq 9.1\times 10^{-2}$.
\begin{figure}[h]
\setlength{\unitlength}{1cm}
\begin{center}
\hspace*{-1.6cm}
\epsfig{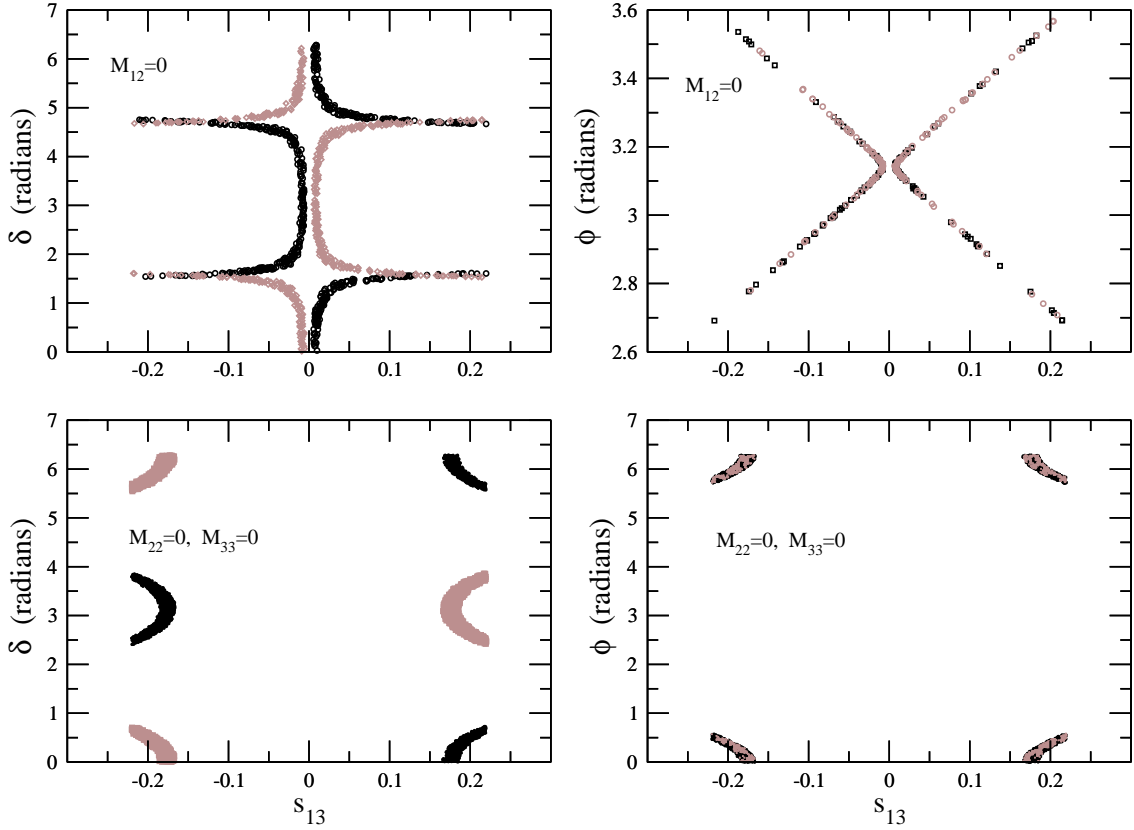}
\end{center}
\caption{ \it The same as fig.2 for the three possible texture zeros in IH. As for NH no
solution exists with $\theta_{13}=0$.}
\label{fig3}
\end{figure}
The solution in terms of the parameter spaces 
($s_{13},\phi$) and ($s_{13},\delta$) is shown in fig.2. 
Notice that since 
$\frac{\epsilon}{m}\simeq 6\times 10^{-2}$, it follows from eqs.(59),(60) 
that $s_{13}$ depends quite weakly on the phases 
$\phi$, $\delta$ as can 
\vspace{0.2cm}
be seen in fig.2. \\
(d) $M_{22}=0$ \\
This case leads to 
\be
\frac{m}{2}+\epsilon \left(\pm 1 \pm \frac{\gamma_{31}}{\sqrt{2}}\right)^2 e^{i\phi}=0
\ee
\vspace{0.2cm}
which is clearly impossible to satisfy. \\
(e), (f) $M_{23}=0$, $M_{33}=0$ \\
Comparison of these two matrix elements with $M_{22}$, as seen from eqs.(54), leads
to the immediate conclusion that these conditions can also not be satisfied. Hence:

\newpage
{\it Conclusion: two texture zeros are possible for $\theta_{13}\ne 0$ in NH case,
namely $M_{12}=0$ and $M_{13}=0$ with $8.1\times 10^{-2}\leq |s_{13}| \leq 9.1\times 10^{-2}$.}

\vspace{0.3cm}

We now turn to texture zeros in IH. In contrast with NH, they cannot be studied 
analytically in a straightforward way with the exception of $M_{11}=0$. We have
in this case
\be
\tilde me^{i\phi}=2\sqrt{\Delta m^2_{\otimes}-\Delta m^2_{\odot}}
\ee
which is clearly impossible to satisfy (recall that $\tilde m=\sqrt{\Delta m^2_{\otimes}}$).
All other 5 cases were investigated numerically. We found three possible 
texture zeros: $M_{12}=0,M_{22}=0,M_{33}=0$. They are displayed in fig.3 in the parameter
spaces ($s_{13},\phi$) and ($s_{13},\delta$). Owing to the structure of these three
matrix elements, it is readily seen that two possible solutions exist for each texture.
They correspond to the black and grey areas of fig.3. $M_{12}=0$ implies 
$|s_{13}|\geq 7\times 10^{-3}$ whereas $M_{22}=0,M_{33}=0$ are excluded up to $2\sigma$,
as $|s_{13}|\geq 0.18$. It is seen that no solution exists
for $\theta_{13}=0$ as derived in sec.3. 

\vspace{0.3cm}
{\it Conclusion: three texture zeros are possible for $\theta_{13}\ne 0$ in IH case,
namely $M_{12}=0,M_{22}=0,M_{33}=0$. The first implies $|s_{13}|\geq 7\times 10^{3}$ and
the second and third imply $|s_{13}|\geq 0.18$.}

\vspace{0.3cm}

As far as equalities between matrix elements of M are concerned and 
since only 6 of these elements are independent, one is lead to 15 possible equalities
for each hierarchy. Again, we require $|s_{13}|\leq 0.16$ (90\% CL) or $|s_{13}|\leq 0.22$ 
(99.73\% CL) and organize the analysis considering the pairs with $M_{11},M_{12},M_{13},M_{22},
M_{23}$ as follows
\begin{center}
\begin{tabular}{cccccccccccccc}
$M_{11}$&$M_{12}$& &$M_{12}$&$M_{13}$& &$M_{13}$&$M_{22}$& &$M_{22}$&$M_{23}$& &$M_{23}$&$M_{33}$ \\
      &$M_{13}$& &      &$M_{22}$& &      &$M_{23}$& &      &$M_{33}$& &      &       \\
      &$M_{22}$& &      &$M_{23}$& &      &$M_{33}$& &      &      & &      &       \\
      &$M_{23}$& &      &$M_{33}$& &      &      & &      &      & &      &       \\
      &$M_{33}$& &      &      & &      &      & &      &      & &      &       \\
\end{tabular}
\end{center}
So taking first $M_{11}=M_{12}$ (see eq.(54)) and using $\alpha_{22},\alpha_{23},
\beta_{12},\beta_{22},\gamma_{13}$, as in eqs.(5), (10), (11) we have \\
\be
s_{13}e^{2i\delta}m=\frac{1}{\sqrt{2}}e^{i\delta}(\pm \epsilon e^{i(\phi-2\delta)}\pm m),
\ee
\be
{\rm or~alternatively}~~~~~~~~\epsilon e^{i\phi}+s_{13}^{2}e^{2i\delta}m=-\epsilon e^{i\phi}
\pm\frac{1}{\sqrt{2}}s_{13}e^{-i\delta}\epsilon e^{i\phi}\pm s_{13}e^{i\delta}m
\ee
where we eliminated the solution $s_{13}=0$ to obtain the first of these. Equating moduli
in (63) 
\be
s_{13}m=\frac{m}{2}\left(1\pm \frac{\epsilon}{m}cos(\phi-2\delta)+{\rm O}(\frac
{\epsilon^{2}}{m^2})\right)
\ee
which is clearly impossible to satisfy, owing to the experimental constraints on $s_{13},m$ and 
$\epsilon$, as seen above. Eq.(64) is also impossible to satisfy as can be numerically checked. 
Hence we conclude that $M_{11}=M_{12}$ can only be verified for $s_{13}=0$ and with an overall
positive sign in front 
in the term $3\epsilon\alpha_{22}\beta_{22}\beta_{12}e^{i\phi}$ in $M_{12}$. In this case we 
recover the form of $M_{11}$ and $M_{12}$ as in eqs.(15), (17). The same arguments can be used
to prove that $M_{11}=M_{13}$ is also impossible unless $s_{13}=0$, since $M_{13}$ is 
identical to $M_{12}$ except for the replacement $\alpha_{23},\alpha_{22} \rightarrow 
\alpha_{33},\alpha_{32}$ (see eq.(54)). It is also aparent that $M_{11}=M_{22}$, $M_{11}=M_{23}$ 
and $M_{11}=M_{33}$ cannot be satisfied because whereas the coefficients of $\epsilon
(=\frac{1}{3}\sqrt{\Delta m^2_{\odot}})$ are of the same order in all four matrix entries, those 
of $m(=\sqrt{\Delta m^2_{\otimes}})$ in $M_{11}$ differ by at least one order of magnitude
from the corresponding ones in $M_{22},M_{23},M_{33}$. (A similar argument would hold for the 
above comparison between $M_{11},M_{12}$). 
Considering next $M_{12}$, the comparison between $M_{12}$ and $M_{13}$ (54) shows that although 
$\alpha_{23}$ and $\alpha_{33}$ may be equal, orthogonality of the matrix $U_{23}$ implies the 
relative signs of the terms of $\epsilon e^{i\phi}$ coefficients to be different in the 
two entries. So one cannot have $M_{12}=M_{13}$ unless $s_{13}=0$.

Use of the above arguments shows that none of the remaining 8 equalities between mass
matrix elements can be satisfied for finite $\theta_{13}$, a result that can as well be
numerically checked.

We finally refer to the equalities in IH. As before, examining first the five possible cases 
involving $M_{11}$ (see eq.(55)), and since the magnitudes of $\beta_{11}(=\sqrt\frac{2}{3}),
\beta_{12}(=\sqrt\frac{1}{3})$ are quite different from all combinations of $\alpha^{'}s$ and
$\beta^{'}s$ that multiply $\tilde m-\lambda$ and $\tilde m e^{i(\phi-\psi)}$, it follows that all
five such equalities are impossible to satisfy. Furthermore, as for NH, the orthogonality of 
the matrix $U_{23}$ prevents all remaining 10 equalities involving 
$M_{12},M_{13},M_{22},M_{23},M_{33}$ unless $s_{13}=0$ in some cases. We are thus lead to the 
following:
 
\vspace{0.4cm}
{\it Conclusion: for both NH and IH there are no possible equalities between matrix elements 
with $\theta_{13}\ne 0$. If $\theta_{13}=0$ some equalities become possible which are the 
obvious ones encountered before.}

\section{Summary}

We have investigated the prospects for neutrino mass matrices with vanishing determinant 
for vanishing and finite
$\theta_{13}$. The vanishing determinant condition alone is expressed by two real conditions,
so the original nine independent parameters in these matrices are reduced to seven. Hence the
undesirable situation of existing and planned experiments not being able to determine all 
these nine quantities is in this case overcome. Furthermore, as shown in the introduction, 
the vanishing of $\theta_{13}$ implies that the CP violating Dirac phase is unobservable 
and the mass matrix can be diagonalized by a real and orthogonal matrix.
In such case the mass matrix determinant is basis independent, $det~M=det~D$, while the
vanishing determinant condition is always basis independent. So $det~M=0$ is always 
equivalent to the lightest neutrino being massless. On the other hand, if $\theta_{13}\ne 0$
one has in general $det~M\ne det~D$, while the vanishing determinant condition remains
basis independent.

We considered both the normal and inverse mass hierarchies.
Summarizing our main conclusions for vanishing determinant mass matrices

{\it(i) $\theta_{13}=0$} 

In the case of normal hierarchy there can be no observable  
$\beta\beta_{0\nu\nu}$ decay. For inverse hierarchy 
$\beta\beta_{0\nu\nu}$ decay depends on the Majorana phase and can be observed in the 
next generation of experiments for all or most of the possible phase range.
Texture zeros and equalities between mass matrix elements besides the obvious ones are 
incompatible with experimental evidence.

{\it(ii) $\theta_{13}\ne 0$} 

$\beta\beta_{0\nu\nu}$ decay satisfies the same properties as for $\theta_{13}=0$ in
both normal and inverse hierarchies whereas texture zeros become possible in this case. 
In NH for $\theta_{13}\ne 0$, one may have $M_{12}=0$ and $M_{13}=0$ if 
$8.1\times 10^{-2}\leq |s_{13}|\leq 9.1\times 10^{-2}$ (fig.2) and in IH 
$M_{12}=0$ if $|s_{13}|\geq 7\times 10^{-3}$ 
(fig.3). Also in IH $M_{22}=0,M_{33}=0$ are possible but with rather large $s_{13}$,
namely $|s_{13}|\geq 0.18$ which is excluded at $2\sigma$ (fig.3). All equalities between 
mass matrix elements both in NH and IH are those which reduce to the obvious ones in the 
limit $\theta_{13}=0$: there are no such equalities if $\theta_{13}\ne 0$.

\vspace{0.4cm}
\noindent {\Large \bf Acknowledgements}

{\em MP wishes to thank Centro de F\'{\i}sica Te\'{o}rica das Part\'{\i}culas for
hospitality and financial support.
The work of BCC was supported by Funda\c{c}\~{a}o para a
Ci\^{e}ncia e a Tecnologia through the grant SFRH/BPD/5719/2001.}

\end{document}